# DATA MINING TECHNIQUES: A SOURCE FOR CONSUMER BEHAVIOR ANALYSIS

Abhijit Raorane[1] &  R.V.Kulkarni[2]

[1]Department of computer science, Vivekanand College, Tarabai park Kolhapur
abhiraorane@gmail.com
[2]Head of the Department, Chh. Shahu Institute of business Education and Research Centre Kolhapur. 416006 drrvkulkarni@siberindia.co.in

**Abstract**

*Various studies on consumer purchasing behaviors have been presented and used in real problems. Data mining techniques are expected to be a more effective tool for analyzing consumer behaviors. However, the data mining method has disadvantages as well as advantages. Therefore, it is important to select appropriate techniques to mine databases. The objective of this paper is to know consumer behavior, his psychological condition at the time of purchase and how suitable data mining method apply to improve conventional method. Moreover, in an experiment, association rule is employed to mine rules for trusted customers using sales data in a super market industry*

1.   **Introduction**

The study of the person who wants to buy certain product from shopping place, at the time of purchasing product, why he wants to buy it ? This is very interesting concept to study. To study his psychological mindset and converting this into statistical format and see that is there any technical format by which we can analyze his buying behavior.

The study of consumers helps firms and organizations to improve their marketing strategies by understanding issues such as how

1)   The psychology of consumers that how he thinks, feel, reasons and select between different alternatives.

2) The mindset of how the consumer is influenced by his or her environment.
3) The behavior of consumer while shopping or making other marketing decisions.
4) How customer motivation and decision strategy differ between products that differ in their level of importance or interest that they entail for the customer; and
5) How management can adjust and improve their marketing campaigns and marketing ideas to more effectively reach customer.

**1.1  Consumer behavior**

Consumer behavior means the study of individuals, groups or organizations about their process of selecting, securing, using and disposing the products, services, experiences or ideas to satisfy needs and the impact of these process on the consumer and the society.

Behavior concerns either with the individual or the group (e.g. In college friends influence what kind of clothes a person should wants to wears) or a firm (peoples working in firm make decision as to which products the firm should use.) The use of product is often so important to the marketer because this may influence how a product is best positioned or how we can encourage increased consumption.

Consumer behavior involves services and ideas as well as tangible products.

**1.2  Application of consumer behavior**

The application of consumer behavior is widely acknowledged in making of marketing strategy i.e. for making better marketing campaigns. For example by understanding that children's are more receptive to food advertising i.e. when they are watching the cartoon films after their school timing, they are also hungry, so we learn to schedule snack advertisements late in the afternoon.

A second application is public policy. There are many deaths occurred due to chewing of tobacco products as well as smoking. So the government i.e. Federal drug administration (FDA) made it compulsory to tobacco manufacturing companies to print the warning message on the tobacco packets, pouch and containers. Social marketing involves getting ideas across to consumers rather than selling something.

As a final benefit, studying consumer behavior also correlate consumers common sense, e.g. if you buy a one litter bottle of ketchup for 65 Rs, at the same time you pay higher, when you purchase half litter bottle. How ever if you often pay a size premium by buying the larger quantity. In other words in this case knowing this facts will sensitize you to the need to check the unit cost labels to determine if you are really getting a bargain.[1,2]

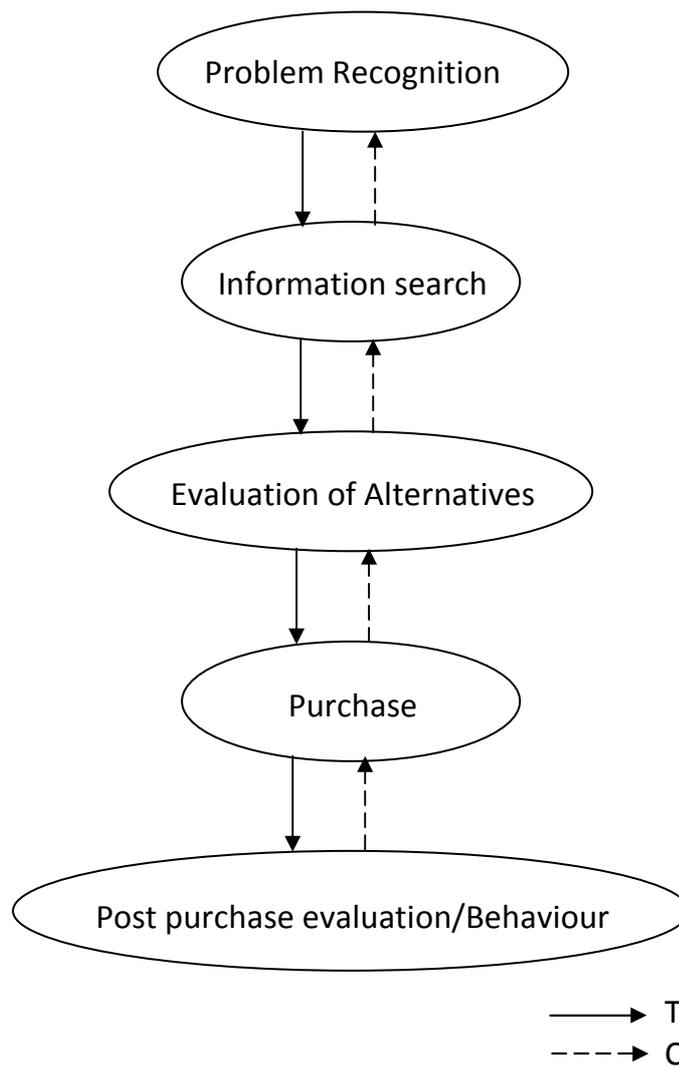

In "Market basket analysis in multiple store environments" the author Yen-ling chen, Kwei Tang, Ren-Jie shen, Ya-han Hu find out that there are two main problem in using the existing methods which are used in a multi-store environment. The first is caused by the temporal nature of purchasing pattern. An apparent example is seasonal products. The second problem is associated with finding common association pattern in subset of store. To overcome this problem the authors develop an Apriori like algorithm for automatically extracting association rules in multi-store environment. [3]

Various studies on consumer purchasing behavior have been presented and used in real Problem. Data mining techniques are expected to be more effective tool for analyzing consumer behavior. However the data mining methods has disadvantages as well as advantages. Therefore it is important to select appropriate tool to mine database. The Junzo watada and kozo yamashiro in their paper "A Data mining approach to consumer behavior" tried to improve data mining analysis by applying several methods including fuzzy clustering, principal component analysis and discriminate analysis. Many defects included in the conventional methods are improved in this paper. [4]

S. vijaylaxmi, V, Mohan, S. Suresh Raju in their paper " Mining of users' access behavior for frequent sequential pattern from web logs" explains, In sequential pattern Mining comes in association rule mining. For a given transaction database T, an association rule is an expression of form X→Y holds with confidence τ % of transaction set T if σ % of transaction set T support X U Y. Association rule Mining can be divided in to two steps. Firstly frequent pattern with respect to support threshold minimum support are mined. Secondly association rules are generated with respect to confidence threshold minimum confidence. [5]

Parvinder S. Sandhu, Dalvinder S. Dhaliwal and S. N. Panda in paper "Mining utility-oriented association rules" explains, An efficient approach based on profit and quantity" Association rule mining has been an area of active research in the field of knowledge discovery and numerous algorithms have been developed to this end. Of late, data mining researchers have improved upon the quality of association rule mining for business development by incorporating the influential factors like value (utility), quantity of items sold (weight) and more, for the mining of association patterns. In this paper, they propose an efficient approach based on weight factor and utility for effectual mining of significant association rules. Initially, the proposed approach makes use of the traditional Apriori algorithm to generate a set of association rules from a database. The proposed approach exploits the anti-monotone property of the Apriori algorithm, which states that for a k-itemset to be frequent all (k-1) subsets of this itemset also have to be frequent. Subsequently, the set of association rules mined are subjected to weightage (W-gain) and utility (U-gain) constraints, and for every association rule mined, a combined utility weighted score (UW-Score) is computed. Ultimately, they determine a subset of valuable association rules based on the UW-Score computed. The experimental results demonstrate the effectiveness of the proposed approach in generating high utility association rules that can be lucratively applied for business development. [6]

      B. Yıldız and B. Ergenç (Turkey) in "Comparison of Two Association Rule Mining Algorithms without Candidate Generation" Association rule mining techniques play an important role in data mining research where the aim is to find interesting correlations among sets of items in databases. Although the Apriori algorithm of association rule mining is the one that boosted data mining research, it has a bottleneck in its candidate generation phase that requires multiple passes over the source data. FP-Growth and Matrix Apriori are two algorithms that overcome that bottleneck by keeping the frequent itemsets in compact data structures, eliminating the need of candidate generation. To their knowledge, there is no work to compare those two similar algorithms focusing on their performances in different phases of execution. In this study, they compare Matrix Apriori and FP-Growth algorithms. Two case studies analyzing the algorithms are carried out phase by phase using two synthetic datasets generated in order i) to see their performance with datasets having different characteristics, ii) to understand the causes of performance differences in different phases. Their findings are i) performances of algorithms are related to the characteristics of the given dataset and threshold value, ii) Matrix Apriori outperforms FP-Growth in total performance for threshold values below 10%, iii) although building matrix data structure has higher cost, finding itemsets is faster. [7]

      "EXTRACTION OF INTERESTING ASSOCIATION RULES USING GENETIC ALGORITHMS" Peter P. Wakabi-Waiswa* Venansius Baryamureeba, In this paper they describes the process of discovering interesting and unexpected rules from large data sets is known as association rule mining. The typical approach is to make strong simplifying assumptions about the form of the rules, and limit the measure of rule quality to simple properties such as support or confidence. Support and confidence limit the level of interestingness of the generated rules. Comprehensibility, J-Measure and predictive accuracy are metrics that can be used together to find interesting

association rules. Because these measures have to be used differently as measures of the quality of the rule, they can be considered as different objectives of the association rule mining problem. The association rule mining problem, therefore, can be modelled as a multi-objective problem rather than as a single-objective problem. In this paper we present a Pareto-based multiobjective evolutionary algorithm rule mining method based on genetic algorithms. Predictive accuracy, comprehensibility and interestingness are used as different objectives of the association rule mining problem. Specific mechanisms for mutations and crossover operators together with elitism have been designed to extract interesting rules from a transaction database. Empirical results of experiments carried out indicate high predictive accuracy of the rules generated. [8]

Jyothi Pillai in "User centric approach to itemset utility mining in Market Basket Analysis" describes Business intelligence is information about a company's past performance that is used to help predict the company's future performance. It can reveal emerging trends from which the company might profit . Data mining allows users to sift through the enormous amount of information available in data warehouses; it is from this sifting process that business intelligence gems may be found. Within the area of data mining, the problem of deriving associations from data has received a great deal of attention. This problem is referred as "market-basket problem". Association Rule Mining (ARM), a well-studied technique in the data mining field, identifies frequent itemsets from databases and generates association rules by assuming that all items have the same significance and frequency of occurrence in a record. However, items are actually different in many aspects in a number of real applications such as retail marketing, nutritional pattern mining, etc. Rare items are less frequent items. For many real world applications, however, utility of rare itemsets based on cost, profit or revenue is of importance. For extracting rare itemsets, the equal frequency based approaches like Apriori approach suffer from "rare item problem dilemma". Utility mining aims at identifying rare itemsets with high utility. The main objective of Utility Mining is to identify the itemsets with highest utilities, by considering profit, quantity, cost or other user preferences. Also valuable patterns cannot be discovered by traditional non-temporal data mining approaches that treat all the data as one large segment, with no attention paid to utilizing the time information of transactions. Now, as increasingly complex real-world problems are addressed, temporal rare itemset utility problem, are taking center stage. In many real-life applications, high-utility itemsets consist of rare items. Rare itemsets provide useful information in different decision-making domains such as business transactions, medical, security, fraudulent transactions, and retail communities. For example, in a supermarket, customers purchase microwave ovens or frying pans rarely as compared to bread, washing powder, soap. But the former transactions yield more profit for the supermarket. A retail business may be interested in identifying its most valuable customers i.e. who contribute a major fraction of overall company profit. In this paper, these problems of analyzing market-basket data are considered and important contributions are presented. It is assumed that the utilities of itemsets may differ and determine the high utility itemsets based on both internal (transaction) and external utilities. [9]

"Efficient Association Rule Mining for Market Basket Analysis" Shrivastava A., Sahu R. writes in that Data mining is an attitude that business actions should be based on learning, that informed decisions are better than uninformed decisions, and that measuring results is beneficial to the business. Data mining is also a process and a methodology for applying the tools and techniques. Association rule mining is also one among the most commonly used techniques in Data mining. A typical and the most running example of association rule mining is market basket analysis. This process analyzes customer buying habits by finding associations between the different items that customers place in their "shopping baskets". The discovery of such associations can help retailers develop marketing strategies by gaining insight into which items are frequently purchased together by customer and which items bring them better profits when placed with in close proximity. The algorithms for single dimensional association rule mining, such as apriori and the FP-tree developed are in a greater use today. However, candidate set generation in apriori is still costly, especially when there exists a large number of patterns and/or long patterns. And both these algorithms prune the itemsets based on their frequencies (i.e., if their frequencies exceed minimum support threshold then they term them as frequent and the rest of them as infrequent). But this pruning technique is insufficient to help market analyst to make decisions such as planning the supermarket's shelf space, changing the layout new store layouts, new product assortments, which products to put on promotion so as to improve their marketing profits. So the focus of this paper is to enhance these algorithms in a way that it provides frequent profitable patterns which help market analyst to make the best informed decisions for improving their business. [10]

**2. Methodology for analysis of consumer behavior**

Over the years Data mining (DM) can used to understand the consumer buying behavior using various techniques. Data mining has gradually increases many folds and today it is a giant 100 billion dollar industry. In data mining world every activity of a consumer in a supermarket is treated as a byte of data. How the consumer spends, which day what time normally he/she does the shopping, what they buy most often, how much they buy, in that locality etc. All this data which is gathered somewhere at the backend about which a consumer is not even aware and there is a big industry which is slicing & dicing this data & selling it at a premium price. [11]

**2.1 What is data mining?**

Data mining is the method of analyzing data from different angle or perspective and collecting it to get useful information that can be used to increase revenue costs or both, DM allows backend processors to analyze data from many different dimensions, categories it & summarize the relationships identified.

Technically, data mining is the process of finding correlations or patterns among dozen of fields in large relational databases .Data mining is primarily used day by day comprise with a strong consumer focus retail, financial , communication & marketing organizations. It enables this companies to determine relationships among internal factors such as price, product positioning or staff skills & external factors such as economic indicators, competition & customer demographics.

Most of the time the data is used to analyze the pattern or shopping habits of consumers like in festive season which product sells more. What are the association between these product? Is there any pattern in this habit? If data show some common theme then stores management arrange that product accordingly. e.g. If management arrange electronic product like television, LCDs, Tape recorders, Mobiles etc. with attractive schemes in the front row in festive season. And also arrange the similar items which customer tends to buy along with these product. To make more profit stores will not run any discount or special offers on the products on busy days. Yet another area commonly traced is the weekly shopping habit of the customer, What products they buy and of what quality. This information can be used for stocking purposes and handle the inventory cost. Likewise there are many other aspects in which this data analysis is leading to better consumer satisfaction. For monthly analysis about the certain product demand i.e. buying in the start of the month and buying at the end of the month? The people have money to spent in the starting of the month and at the end of the month people spend less. In the vacation of the school and at the starting of the school the requirement of certain commodity is increase. So to maintain the inventory and also to increase sell in this period. It is necessary to grab this opportunity of consumer needs.

In all these cases the data which is collected from different sources. Some of these are operational or transactional data such as sales, cost, inventory, payroll and accounting.

Non operational data such as industry sales, forecast data and macro economic data. Meta-data Data about data itself, such as logical database design or data dictionary definition. [12]

## 2.2 Classification of Data mining system
### 2.2.1 Categorization of the data mining system according to the kind of knowledge mined -

Data mining system can be categorized according to the kinds of knowledge they mine, that is based on functions of data mining, such as characterization. Discrimination, association, classification, clustering, outlier analysis and evaluation analysis. A comprehensive data mining system usually provides multiple integrated data mining functionalities.

Moreover data mining system can also be differentiated based on the levels of knowledge abstraction, including generalized knowledge (at a raw data level) or knowledge at multiple levels (considering several levels of abstraction). An advanced data mining system should facilitate the discovery of knowledge of multiple levels of abstraction.

Data mining system can also be categorized as those that mine data regularities(commonly occurring patterns) versus those that mine data irregularities(such as exception, or outliers). In general concept description, association analysis, classification, prediction and clustering mine data regularities, rejecting outlier as noise. These methods may also help to detect outliers.

**2.2.2  Classification according to the kinds of techniques utilized –**

Data mining system can also be classified according to the techniques adopted or employed. These techniques can be described according to the degree of user interaction involved ( e.g. autonomous system, interactive exploratory system, query- driven system ) or the method of data analysis employed ( e.g. Database – oriented or data warehouse oriented techniques, machine learning statistics, visualization, pattern recognition neural  network and so on ) A sophisticated data mining system will often adopt multiple data mining techniques or work out a effective, integrated technique that combine the merits of a raw individual approaches.[13,22]

**2.3    Classification according to the application adopted**

Data mining systems can also be classified according to the application they used for. for example, there could be data mining system developed specifically for telecommunications, DNA, stocks markets, web sites, e-mail, and so on. Different application often required the integration of application–specifics method .Therefore a generic, all purpose data mining system may not fit domain– specific mining task. [13]

**3.     Research Methodology**

For actual testing & getting the result by implementing new methodologies in data mining, the researcher gone through all the details about the consumer behavior & he experiment it by choosing a organization a mall or super market as a sample for his study. Researcher collect all the live data day wise, month wise i.e. transaction of each customer.

After collecting the data researcher searches various methodologies which go through the methodology for finding the answer. He choose the suitable technique, formula, algorithms, methods for the customer data base.

Researcher choose Janata Bazzar , a super market in Kolhapur city for his study. Collects all the customer buying transaction database.

After collecting the data, researcher select the suitable method from various alternatives. He select association rule for checking the association between the various products which are bought by the customer. He implement the market Basket Analysis for this database.[14,22,23]

**4. Actual use of Data Mining Technique- with study unit.**

**4.1 Association rule of mining**

Association rule mining finds interesting association or correlation relationships among a large set of data items, the researcher is become more interested in mining association rules from organizations database . The discovery of interesting association relationship among huge amounts of customer transaction records can help in many business decision making processes such as catalog design , cross marketing and loss-leader analysis.[15,17,18]

**4.2 Transaction of customer with the study unit**

A typical process, the researcher finds from the association rule mining is market basket analysis, This process analyzes customer buying habits by finding association between different items that customers places their "shopping basket". The discovery of such association can help retailers develop marking strategies by gaining insight into which items are frequently purchased together by customers. For instance; if customers are buying milk, now likely are they to also buy bread  (and what kind of bread) on the same trip to the supermarket? Such information can lead to increased sales by helping retailers to selective marketing and plan their shelf space. For example placing milk and bread within close proximity may further encourage the sale of these items together within single visit to the store.

Researcher selects Market Basket Analysis for his data analysis because Market Basket analysis is a tool of knowledge discovery about co-occurrence of nominal or categorical items. Market Basket Transaction or market Basket Analysis is a data mining technique to derive association between data sets. Researcher has categorical data of transaction records as input to the analysis and the output of the analysis is association rules as a new knowledge directly from data. [16,17]

Researcher have following transactional data from an organization and the numbers of transactions in one day are limited as the data shown below.

| Transaction ID | Items from the customer who bought more than one item |
|---|---|
| 1 | Sugar, wheat, pulses, Rice |
| 2 | Sugar, pulses |
| 3 | Wheat, pulses |
| 4 | Pulses, wheat, Rice |
| 5 | Wheat, pulses |
| 6 | Sugar, Wheat |
| 7 | Sugar, Rice, pulses |

Based on the data above, Researcher derive the following output of association rule by using market Basket analysis.

**Output- Association Rules**

| People who bought this item | Also bought the following items | Support | Confidence |
|---|---|---|---|
| Wheat | Pulses | 57% | 80% |
| Rice | Pulses | 43% | 100% |

The association rule will have the following form

$$X \rightarrow Y$$

that form has meaning that people who bought items of set x are often also bought items on set Y e.g if X = { sugar, wheat} and Y ={Rice, Pulses} and we get association rule indicates that people who bought sugar and wheat also bought Rice and pulses.

<u>Support and confidence are two measures of association rules</u>. Support is the frequency of transaction to have all the items on both sets and Y are bought together. For e.g. a support of 5% shows that percentage of all transaction (that researcher consider for the analysis) indicates that items on set X and Y are purchased together. In formula, support can be computed as probability of the union set X & Y

$$\text{Support } (X \rightarrow Y) = P(X \cup Y) = \frac{n(X \cup Y)}{N}$$

Notation of support count indicates that the total frequency of the set union and is the total number of transaction for the analysis. A rule that has very low support may occurs simply by chance. We can also view support as the numbers of instances that the association rules will predict correctly.

Confidence of 80% shows that 80% of the customer who bought items on set X also bought items on set Y. In formula, confidence is computed as conditional probability to obtain set Y given set X. The conditional probability also can be computed through proportion of support.

$$\text{Confidence}(X \to Y) = P(X/Y) = \frac{n(X \cup Y)}{n(X)}$$

Notation is total frequency of set X. Confidence is a measures of accuracy or reliability about the inference made by the rule that the number of instances that the association rules will predict correctly among all instance it applies to.

To obtain the association rules Researcher usually apply two criteria:
1. minimum support
2. minimum confidence [19,20,21]

**Conclusion**

1) Data Mining System is useful to study buying behavior of the customers in retail departmental stores. With this study researcher has concluded that there are certain buying habits of the customers. And according to this buying habits of customer, management may update their system of providing various types of services to their customers to delight the customers and to retain the customer with same business house.

2) The data mining system is useful to Business house to find out the association of the customers with different products. And how customers are shifting from one brand to another brand of product to satisfy their need because their earlier buying habits are properly studied by the Data mining System.

**Recommendation**

The knowledge generated from Data mining technique is useful for the organizations engaged in Retailing business for their decision making process.